\documentclass[doublecol,cite]{epl2}

\usepackage{graphicx,epsfig}
\usepackage{subfigure}  
\usepackage{amsmath}
\usepackage{amsfonts}
\usepackage{xfrac}
\usepackage{enumitem}
\usepackage{xcolor}
\usepackage{color}
\usepackage{epstopdf}
\usepackage{ulem}
\usepackage[hidelinks]{hyperref}
\usepackage[capitalise]{cleveref}
\usepackage{bbold}
\usepackage{fancyhdr}

\title{Emergent States in Systems of Chiral Self-Propelled Rods}
\date{February 9, 2023}

\author{Rüdiger Kürsten\inst{1,2,3} and Demian Levis \inst{1,2}}
\shortauthor{B. Kürsten, D. Levis}

\institute{                    
  \inst{1} Departament de F\'isica de la Mat\`eria Condensada, Universitat de Barcelona - C. Mart\'i Franqu\`es 1, 08028 Barcelona, Spain.\\
  \inst{2} UBICS University of Barcelona Institute of Complex Systems - C. Mart\'i i Franqu\`es 1, E08028 Barcelona, Spain.\\
  \inst{3} Institut für Physik, Universität Greifswald, Felix-Hausdorff-Str. 6, 17489 Greifswald, Germany
}

\abstract{
We study inherently chiral self-propelled particles, self-rotating at a fixed frequency, in two dimensions, subjected to nematic alignment interactions and rotational
noise. By means of both, homogeneous and spatially resolved mean
field kinetic theory, we identify various different flocking states. We confirm
the presence of the predicted phases using agent-based simulations, in particular, an homogeneous nematic phase at low frequencies, followed by a microflock pattern phase at larger frequencies, characterized by finite-size nematic  clusters.  
We emphasize that special care has to be taken within the simulations
in order to avoid artifacts, and present a non-standard simulation
technique in order to avoid them.
}
\pacs{05.40.Jc}{Brownian motion}
\pacs{05.20.Dd}{Kinetic theory}
\pacs{64.60.Cn}{Order-disorder transformations}
\begin{document}
\maketitle

\section{Introduction}

Active systems composed of self-propelled interacting units display a rich variety of non-equilibrium dynamic structures, hardly reachable  in equilibrium conditions \cite{BechingerRev}. A salient example is the emergence of collective motion at different scales, from flocks of birds to colloidal assemblies \cite{cavagna2012, BartoloFlock}. 
From the theoretical viewpoint,  such flocking phenomenon has been studied under the framework of simple models à la Vicsek, which consider point-like self-propelled particles  accommodating their velocity with their neighbourhood \cite{VicsekRev,GinelliRev}. 
More recently, self-rotation, on top of self-propulsion, has been introduced in  archetypal active particle models   \cite{BennoPRL, levis2019, ventejou2021, LiebchenLevisRev}, with the aim of describing circle swimmers, i.e. active particles moving, in two dimensions (2D),  along (noisy) circular trajectories with a given handedness. Circle swimmers are an instance of chiral active matter \cite{LiebchenLevisRev}, such as chiral microtubules \cite{afroze2021monopolar}, curved polymers \cite{denk2016, hannezo2022chiral},  spermatozoa close to walls \cite{riedel2005} or chiral self-propelled colloids~\cite{Kummel2013Circular,zhang2020reconfigurable,alvarez2021reconfigurable}, and similar chiral motion can also be achieved at a larger granular scale \cite{barois2020sorting,arora2021emergent}.

The analysis of simple dry models of chiral active particles, or circle swimmers,  with polar velocity alignment interactions, has revealed that chirality does not modify significantly the flocking transition of Vicsek models, but controls the emergence of different kinds of structures in the symmetry broken phase \cite{BennoPRL}. A dominant interaction in most instances of circle swimmers is likely to be due to anisotropic collisions, due to their elongated shape. This typically gives rise to an effective nematic, rather than polar,  alignment \cite{peruani2006SPR}. Although the alignment of such self-propelled rods \cite{PeruaniRev2020, baskaran2010nonequilibrium} partly relies on excluded volume interactions, a popular simplified picture (keeping the  symmetries of the problem), has been to consider a nematic version of the Vicsek model: point-like particles, polar in terms of their motion, but with nematic velocity alignment \cite{ChatePeruani2008, GinelliPeruani2010}. Here, we adopt this perspective and study the impact of chirality, in the form of self-rotation with a given intrinsic frequency $\omega$, on the collective behaviour of nematic, point-like, self-propelled rods. 
A recent work has also considered nematic Vicsek-type chiral particles, with a distribution of rotation frequencies, to investigate the stability of  ordered states against such quenched disorder \cite{ventejou2021}. Here we shall focus on the mono-frequency case, and the emergence of patterns controlled by the intrinsic frequency $\omega$.          

\begin{figure}[h]
    \centering
  \includegraphics[width=0.22\textwidth]{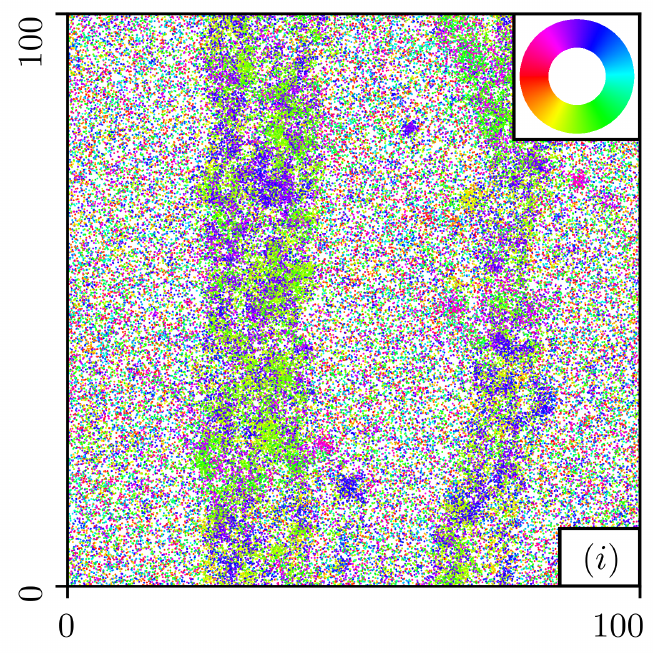}
\includegraphics[width=0.22\textwidth]{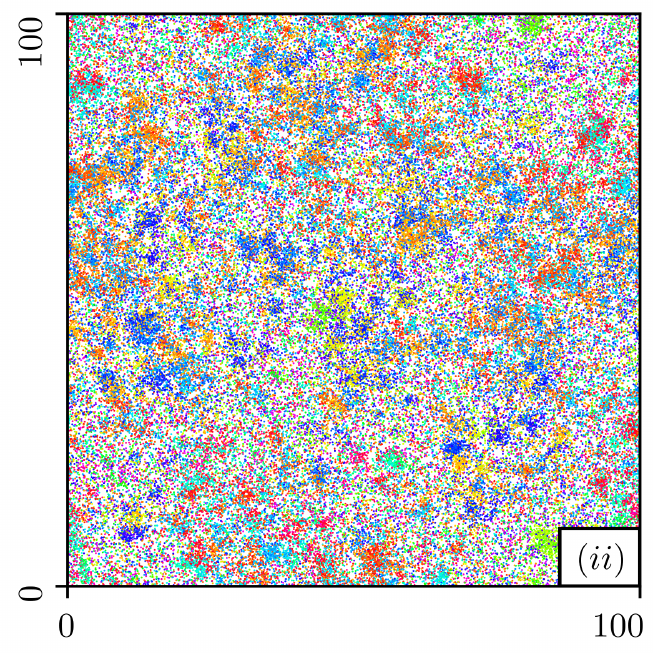}\\
\includegraphics[width=0.22\textwidth]{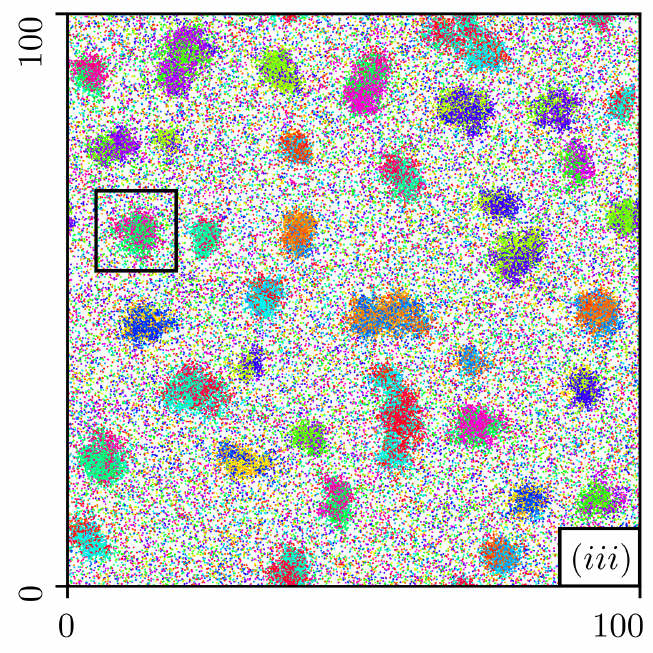}
\includegraphics[width=0.22\textwidth]{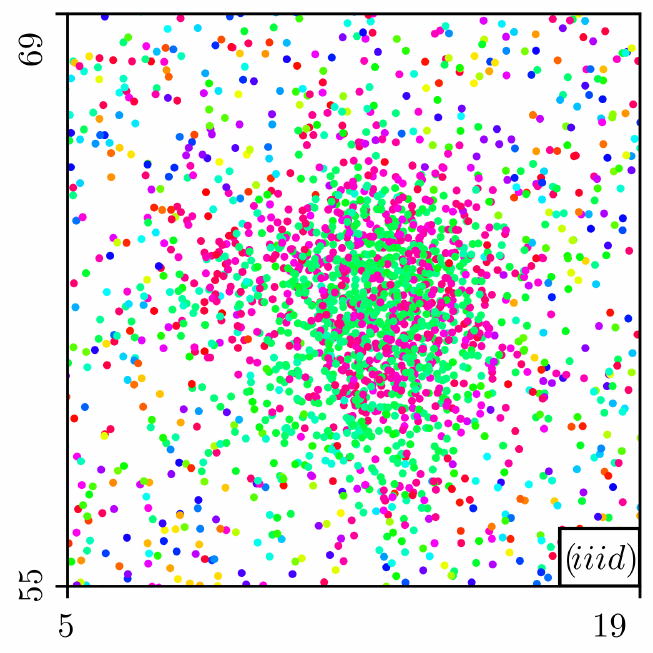}
    \caption{Typical snapshots of the ordered phases from simulations of $N=10^5$ particles,  colored according to their orientation (using the color wheel in $(i)$). 
$(i)$ Non-chiral nematic bands ($\omega=0$, $\Gamma=0.14$). 
$(ii)$ Nematic order, homogeneous on large length scales ($\omega=1$, $\Gamma=0.18$). 
$(iii)$ Nematic microflocks ($\omega=3$, $\Gamma=0.18$).
$(iiid)$ Detailed view of a nematic microflock, marked in $(iii)$ by a black square.
See SM \cite{SM} for numerical details.
}\label{fig:snaps}
\end{figure}

In this Letter, we study the steady phases of  nematically aligning point-like chiral active particles in two-dimensions (2D) by means of both a kinetic theory and particle-based simulations. For efficient simulations, we optimize a non-standard second order discretization scheme for the model under study. The kinetic theory is  derived under a mean-field approximation of the $N$-body Fokker-Planck equation. To grasp the phase behaviour of the system, we  perform a linear stability analysis of the homogeneous solutions of the one-body equations obtained, and then compare our findings with microscopic simulations. 
We find that moderate $\omega$ destabilize nematic bands, yielding a nematic state with no clear patterning but largely homogeneous (see Fig. \ref{fig:snaps}), in a regime where polar interactions would trigger macrophase separation \cite{BennoPRL}.  Larger frequencies induce the formation of rotating microflocks, akin the polar case \cite{BennoPRL}, but featuring nematic order. As we show below, the expectations resulting from the stability analysis, indicating structure formation through a finite wave-length instability for large enough $\omega$, are confirmed by numerical simulations.  

\section{Model}

We consider $N$ self-propelled point-like particles in a $L\times L$ box, with periodic boundary conditions, located at $\bold{r}_i(t)=(x_i(t),y_i(t))$ at time $t$ with orientation $\bold{n}_i(t)=(\cos\phi_i,\sin\phi_i)(t)$. Their motion follows the overdamped dynamics:
\begin{equation}
    \dot{\bold{r}_i}=v\bold{n}_i,\,\ 
    \dot{\phi_i}=\omega+\Gamma\sum_{j\in\partial_i}\sin[2(\phi_j-\phi_i)]+\sqrt{2D}\xi_i
    \label{eq:model_langevin}
\end{equation}
where $\omega$ is the intrinsic frequency of the particles, leading to circle swimming, and $v$ their self-propulsion speed. 
Interactions are described by the sum term, that runs over the nearest neighbours $j$ of particle $i$ defined as $|\bold{r}_i-\bold{r}_j|<R$ (defining our unit of length). The strength of the nematic coupling is quantified by $\Gamma$, while $D$ quantifies the noise strength (also called rotational diffusion, the inverse of which defines our time unit). The noise term $\xi_i$ is Gaussian, white, of zero mean and unit variance.  
In the absence of noise and interactions, particles describe circular trajectories of radius $v/\omega$
at constant velocity.
We simulate systems of sizes up to $N=10^5$  varying  $\Gamma=0..0.3$ and $\omega=0...3$ at fixed number density $\rho=NR^2/L^2=10$, and $v, R, D=1$. This model is an extension of the one introduced in \cite{BennoPRL}, introducing a factor 2 in the interaction term to describe nematic alignment \cite{ElenaPRE}. It differs from the model considered in \cite{ventejou2021} in that the interaction term is not normalized by the number of neighbours. 

\section{Mean-Field Theory\label{sec:meanfield}}

To understand the  phase behavior of the system, we construct a mean-field theory starting from the particle-level equations of motion.
The main assumption relies on factorizing the $N-$body probability $P_N(\{\bold{r}_i, \phi_i\}_{i=1}^N;t)$ of finding the system in a given microstate at time $t$, into $N$ identical one-body probabilities.
Integrating over the degrees of freedom of all but one particle, and performing the thermodynamic limit $N\rightarrow\infty$, we arrive at a nonlinear one-body Fokker-Planck-equation, see e.g. \cite{Frank05}.
We find the long time homogeneous solutions of this equation analytically  and then analyze their linear stability.

The $N$-body Fokker-Planck equation equivalent to the Langevin equation eq. \eqref{eq:model_langevin} reads
\begin{align}
    &\partial_t P_N= \sum_{i=1}^N -\partial_{\phi_i}\bigg\{ \big[\omega +\Gamma \sum_{j=1}^N\theta_{ij}\sin(2(\phi_j-\phi_i)) \big] P_N \bigg\} \notag
    \\
    &+D \partial_{\phi_i}^2 P_N -v\sum_{i=1}^N\bigg\{\cos(\phi_i) \partial_{x_i} + \sin(\phi_i)\partial_{y_i} \bigg\}P_N,
    \label{eq:NFP}
\end{align}
where $\theta_{ij}=1$ if $|\bold{r}_i-\bold{r}_j|\le R$ and 0 otherwise. 
We thus assume 
\begin{align}
    &P_N(x_1, y_1, \phi_1, x_2, y_2, \phi_2, \dots, x_N, y_N, \phi_N)
    \notag
    \\
    &=P_1(x_1, y_1, \phi_1)P_1(x_2, y_2, \phi_2)\dots P_1(x_N, y_N, \phi_N).
    \label{eq:mean_field}
\end{align}

\subsection{Homogeneous States\label{subsec:hom}}

Furthermore, we assume that the one-particle distribution is homogeneous in space, meaning:
\begin{align}
    P_1(x_1, y_1, \phi_1)=p(\phi_1)/L^2.
    \label{eq:homogeneous}
\end{align}
Inserting the {\it{ansatz}} eq.\eqref{eq:mean_field}, \eqref{eq:homogeneous} into eq.\eqref{eq:NFP}, we obtain 
\begin{align}
    &\partial_t p(\phi)=-\partial_{\phi}\bigg\{M \Gamma \big[
    \langle\sin(2\varphi)\rangle \cos(2\phi)
    \notag
    \\
    &-
    \langle\cos(2\varphi)\rangle \sin(2\phi)\big]p(\phi)\bigg\}
    -\omega\partial_{\phi}p(\phi)+D\partial_{\phi}^2p(\phi)
    \label{eq:1FP}
\end{align}
after integrating over all particle spatial coordinates, but one angular one, and taking the limit $N\rightarrow \infty$.
Here $\langle . \rangle$ denotes the angular expectation value with respect to $p(\varphi)$ and $M$ the mean number of neighbors defined as
\begin{align}
    M:=\pi \rho\approx{(N-1)\pi R^2}/{L^2}.
    \label{eq:M}
\end{align}

For simplicity, we start considering the case $\omega=0$.
In a first step we assume that $\langle \sin(2\phi)\rangle$ and $\langle \cos(2\phi)\rangle$ are {\it{a priori}} known. In that case we can easily find the stationary solution of Eq.\eqref{eq:1FP}
\begin{equation}
    p_s(\phi)= \frac{1}{Z}\exp\bigg\{ \frac{M\Gamma \langle \cos[2(\varphi-\phi_0)]\rangle\cos[2(\phi-\phi_0)]}{2D}\bigg\},
    \label{eq:steady_state2}
\end{equation}
with
\begin{equation}
     Z=2\pi I_0\bigg(\frac{M \Gamma}{2D}\langle \cos[2(\varphi-\phi_0)]\rangle\bigg),
    \label{eq:steady_norm2}
\end{equation}
where $I_{\alpha}$ denote modified Bessel functions of the first kind, and for an appropriate choice of $\phi_0$ satisfying
${\sin(2\phi_0)}/{\cos(2\phi_0)}={\langle \sin(2\varphi)\rangle}/{\langle \cos(2\varphi)\rangle}$.
The distribution $p_s$ can thus be interpreted as an equilibrium one with temperature $k_B T=D$ and energy $U(\phi)=-M\Gamma\langle \cos[2(\varphi-\phi_0)]\rangle\cos[2(\phi-\phi_0)]/2$.

We now turn on the general case $\omega \neq 0$. We can write the solution of the angular distribution at long times  as
\begin{align}
    p(\phi, t)=\bigg[2\pi I_0\bigg( \frac{M \Gamma Q}{2D}\bigg)\bigg]^{-1}e^{\frac{M \Gamma Q}{2D}\cos[2(\phi-\phi_0-\omega t)]},
    \label{eq:angular_long_time}
\end{align}
where we defined the nematic order parameter
\begin{align}
    Q:=\langle \cos[2(\varphi-\phi_0-\omega t)]\rangle.
    \label{eq:def_nematic_order}
\end{align}
The long time solution eq.\eqref{eq:angular_long_time} thus depends on $\phi_0$, that determines the orientation of the flocking state, and the nematic order parameter $Q$.
The latter satisfies the following self-consistency equation 
\begin{equation}
    Q=I_1\bigg( \frac{M \Gamma Q}{2D}\bigg)\bigg/ I_0\bigg( \frac{M \Gamma Q}{2D}\bigg),
    \label{eq:self-consitency3}
\end{equation}
One easily checks that the distribution \eqref{eq:angular_long_time} with $Q$ satisfying Eq.~\eqref{eq:self-consitency3} indeed solves Eq. \eqref{eq:1FP}.
Obviously, $Q=0$ is always a solution of Eq. \eqref{eq:self-consitency3}. Depending on the value of $\frac{M\Gamma}{D}$ there can be two more solutions $Q=\pm Q^*$. Both of them describe the same physical state as $Q$ is mapped to $-Q$ when the coordinate system is rotated by $\pi$.
As usual, see e.g. \cite{Frank05, Dawson83},  non-zero solutions exist for couplings above the critical coupling given by the phase transition condition of unit derivative with respect to $Q$ of the right hand side of Eq.~\eqref{eq:self-consitency3} at $Q=0$, namely $M\Gamma=4D$.
Within the spatially homogeneous  theory, the nematic order parameter $Q$ is independent of $\omega$ \footnote{This is consistent with the hydrodynamic description of the polar version of the model developed in \cite{BennoPRL}, up to a factor 2 due to the symmetry of the interactions.}. However, in agent-based simulations we observe that the onset of flocking, as well as the value of the order parameter, depend on $\omega$, cf. Fig. \ref{fig:phasediagram_nemorder} $(a)$. We focus on the formation of spatial structures induced by chirality are studied in the following.

\subsection{Linear Stability Analysis\label{subsec:inhom}}

Here, we no longer assume  $P_1(\mathbf{r}_1, \phi_1)$ to be homogeneous.
We proceed analogously to the homogeneous case  to derive the  one-particle Fokker-Planck Eq.~\eqref{eq:1FP_space_dependent}, cf. SM for details \cite{SM}.

We analyze the linear stability of the homogeneous solution Eq. \eqref{eq:angular_long_time}, similar as e.g. in \cite{BDG06, MBM10, Ihle11}, but here on the level of kinetic theory instead of hydrodynamics.
 We assume a spatial domain $(x, y)\in [0, L]\times [0, L]$ with periodic boundary conditions, and write the deviations from this solution in  Fourier form:
\begin{equation}\label{eq:def_fourier}
    P_1(\bold{r}, \phi)-\frac{1}{L^2}p(\phi, t)= \sum_{klm}F_{klm}(t)e^{ik\phi}e^{i\frac{2\pi}{L}lx} e^{i\frac{2\pi}{L}my}.
\end{equation}
We explicitly calculate the time evolution equation of the vector $\mathbf{F}=\{F_{klm}\}$ of all Fourier modes neglecting quadratic terms, see
SM \cite{SM}.
One can write the time evolution symbolically in operator form
\begin{align}
    \partial_t \bold{F}=\mathbb{A}(t) \bold{F}.
    \label{eq:time_evolution_F}
\end{align}
In the non-chiral case, $\omega=0$, $\mathbb{A}$ is time independent.
Its eigenvalues and eigenvectors can be calculated numerically in order to determine the stability of the homogeneous nematically ordered state.

In the chiral case though, the matrix $\mathbb{A}$ depends explicitly on time due to the time dependence of the homogeneous state.
Thus, also the eigenvectors of $\mathbb{A}$ are time dependent, which introduces some complications in the linear stability analysis.
However, we can make use of the fact that $\mathbb{A}(t)$ is periodic with period
    $T={\pi}/{\omega}$
due to the periodicity of the homogeneous state \eqref{eq:angular_long_time} and its invariance under  $\phi \rightarrow \phi +\pi$.
We now focus on the time evolution of $\bold{F}$ over one period $T$, and denote
\begin{align}
    \bold{F}(T)=\lim_{n\rightarrow \infty, \Delta t=T/n}\bigg[\prod_{s=0}^{n-1}(\mathbb{1}+\Delta t \mathbb{A}(s\Delta t))\bigg] \bold{F}(0),\label{eq:time_evolution_F_T1}
\end{align}
where $\mathbb{1}$ denotes the unit matrix. With the definition
\begin{align}
    \mathbb{A}_T:= \lim_{n\rightarrow \infty, \Delta t=T/n}\bigg[\prod_{s=0}^{n-1}(\mathbb{1}+\Delta t \mathbb{A}(s\Delta t))\bigg]\label{eq:stability_matrix}
\end{align}
eq. \eqref{eq:time_evolution_F_T1} becomes
\begin{align}
    \bold{F}(T)=\mathbb{A}_T\bold{F}(0).
    \label{eq:time_evolution_F_T2}
\end{align}
The advantage of Eq. \eqref{eq:time_evolution_F_T2} over Eq. \eqref{eq:time_evolution_F} is that $\mathbb{A}_T$ does not depend on time.
Hence it can be directly used for a linear stability analysis of the homogeneous nematic state.
In practice, we compute $\mathbb{A}_T$ numerically according to eq. \eqref{eq:stability_matrix} not performing the limit $n \rightarrow \infty$ but using large, but finite $n=10^5$ instead.

\subsection{Numerical Results\label{subsec:numeric}}
We numerically analyze the eigenmodes of $\mathbb{A}$ for $\omega=0$ and $\mathbb{A}_T$ for $\omega>0$.
We use
a system size $L=100$ (corresponding to $N=10^5$) for various values of $\Gamma$ and $\omega$ to allow quantitative comparison with simulations (see below). However, we also checked the consistency of the results using different system sizes.

For the homogeneous disordered state we find no unstable modes in any case.
For $\omega=0$, in the homogeneous nematically ordered phase, we always find spatial instabilities at both, {\it{long wavelength}} $\lambda=L$, and {\it{finite wavelength}}, and thus expect the formation of large scale spatial structures. 
In the chiral case, $\omega>0$, in the homogeneous flocking state we only find a {\it{finite wave length instability at large enough frequencies}} $\omega$ and no instabilities below.

The unstable eigenmodes in angular Fourier space are mainly composed of two components: the largest component for $k=0$ induces mass transport on the spatial length scale given by the spatial wave vectors $l, m$, the other significant component for $k=\pm 2$ induces a change of the nematic order, increasing nematic order in locations of increased density and decreasing nematic order in locations of increased density.
From the linear stability analysis we can not predict the exact character of the steady state in presence of an instability because the long time behavior is determined by nonlinear effects.
However, we expect spatial inhomogeneities at about the length scales given by $l,m$.

We summarize the phase behavior of mean field theory in Fig.~\ref{fig:phasediagram_nemorder}$(b)$.
The non-chiral case $\omega=0$ has an inhomogeneous flocking state $(i)$, characterized by the emergence of nematic bands,  {cf. \cite{GinelliPeruani2010, Peruani16}, within mean field theory it is characterized by a nonzero solution of Eq. \eqref{eq:self-consitency3} and a long wavelength instability of Eq. \eqref{eq:time_evolution_F_T2}.}
For small $\omega$ and large enough coupling, the system is in a homogeneous (nematic) flocking state $(ii)$ {characterized by a nonzero solution of Eq. \eqref{eq:self-consitency3} and no instabilities of Eq. \eqref{eq:time_evolution_F_T2}}.
For larger $\omega$ and large enough coupling {Eq. \eqref{eq:self-consitency3} still has a nonzero solution and Eq. \eqref{eq:time_evolution_F_T2} exhibits a finite wavelength instability that indicates the formation of patterns of finite characteristic size $(iii)$.
From mean-field theory at the linearized level, we can not conclude whether or not this state is nematically ordered.
Finally, for small couplings, Eq. \eqref{eq:self-consitency3} has only the solution $Q=0$, thus the system is disordered $(iv)$.}

\section{Particle-based Simulations\label{sec:simulations}}

\begin{figure}
    \centering
    \includegraphics[width=0.23\textwidth]{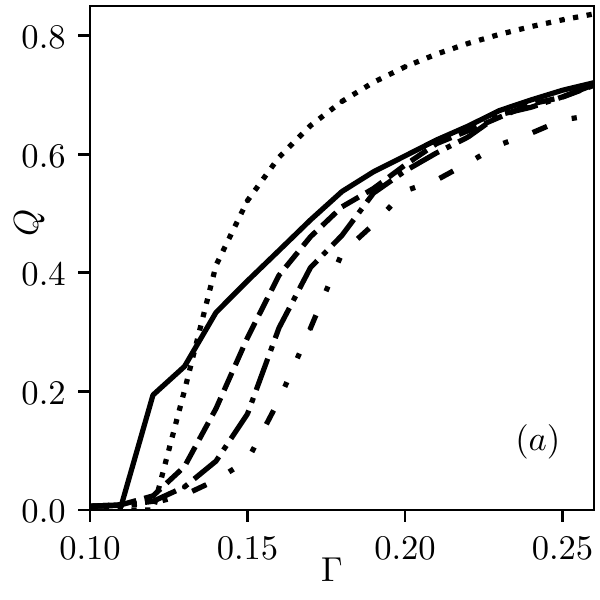}
    \includegraphics[width=0.23\textwidth]{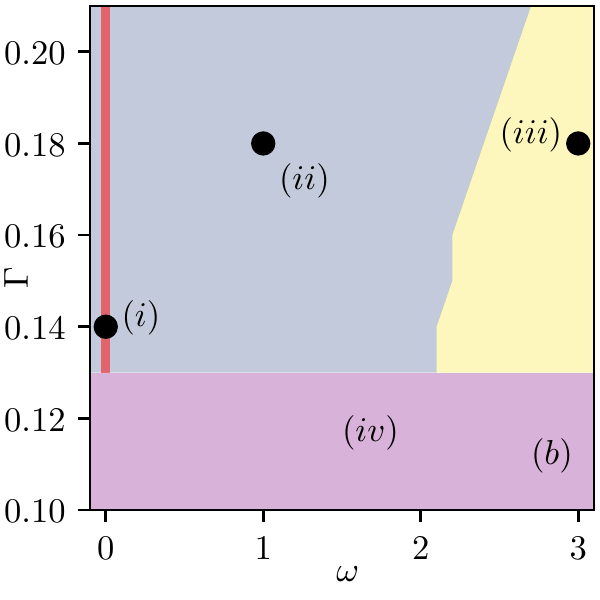}
    \caption{
    $(a)$ Onset of nematic order as a function of coupling for $\omega=0$ (solid), $\omega=0.4$ (dashed), $\omega=1$ (dash-doted), $\omega=2$ (dash-dot-doted) from agent-based simulations. Homogeneous mean field theory is independent of $\omega$ (doted).
    $(b)$ Mean field phase diagram with phases:
    $(i)$ Non-chiral nematic bands. 
    $(ii)$  Nematic order, in mean field homogeneous.
    $(iii)$ Nematic ordered, asynchronous droplets.
    $(iv)$ Disorder.
    Black circles mark parameters of the snapshots shown in Fig. \ref{fig:snaps}.
    See SM \cite{SM} for numerical details.
    }
    \label{fig:phasediagram_nemorder}
\end{figure}

We run agent-based simulations using three different sizes: $N=10^3, 10^4, 10^5$, letting the system evolve to its stationary state from a  uniform and isotropic random initial condition.
Employing the standard Euler-Maruyama scheme with step size $\Delta t=0.01$ within the disordered phase, we observe polar ordered artefacts that are not present when simulating with smaller step sizes.
In order to avoid those artefacts, we employ a non-standard second order integration scheme \cite{KP92}.
In general, it comes at the price of computing interactions with neighbors of neighbors, which is particularly slow at high densities.
For the particular form of the interaction in Eq. \eqref{eq:model_langevin} however, the scheme can be optimized to run at the same complexity as the Euler-Maruyama scheme, cf. SM \cite{SM} for details.
We use the second order scheme for all simulations presented in the letter.
Simulation data are available online \cite{data1,data2}.
The simulations qualitatively confirm the mean field phase diagram, Fig.~\ref{fig:phasediagram_nemorder}$(b)$ as it is discussed in the following in more detail. 

\subsection{Observables}
We measure the polar  and nematic order parameter $P=\frac{1}{N} |\sum_{j=1}^N \exp(\phi_j)|$,  $Q=\frac{1}{N} |\sum_{j=1}^N \exp(2\phi_j)|$,  and the average number of neighbors $\bar{n}=\frac{1}{N} \sum_{j=1}^N \theta_{ij}$. 
We do not find global polar order in any case, meaning $P\approx 0$ in all measurements.
We can distinguish nematic ordered states from disorder by $Q$ and spatially homogeneous states from states with local particle accumulations using $\bar{n}-M$ ($M \approx 31.4$, see Eq.~\eqref{eq:M}).

\subsection{Non-chiral}
For $\omega=0$ we recover the overall phenomenology of similar models of point-like self-propelled particles with nematic, Vicsek-like, alignment  \cite{GinelliPeruani2010}. A flocking transition from a disordered homogeneous state, $(iv)$, to a  state characterized by a net nematic orientation, $(i)$, above some threshold value of the coupling strength, given by $\Gamma_c=4D/M\approx 0.127$ in the mean-field theory. 
The emergence of nematic order is accompanied by a phase separation between a disordered low density gas and high density, nematically ordered lanes, see snapshots Fig. \ref{fig:snaps} $(i)$, which is consistent with the presence of a long wavelength instability of the homogeneous nematic state within mean field theory described above.
In Fig.~\ref{fig:phasediagram_nemorder} $(a)$ we compare the measured nematic order with the homogeneous mean field prediction.
The onset of flocking is shifted towards smaller coupling compared to homogeneous mean field theory.
This shift is most likely caused by phase separation.
Because the density is increased locally within the nematic lanes, nematic order can occur earlier, as a higher density favors order.
\begin{figure*}[h!]
\centering
\includegraphics[width=0.3\textwidth]{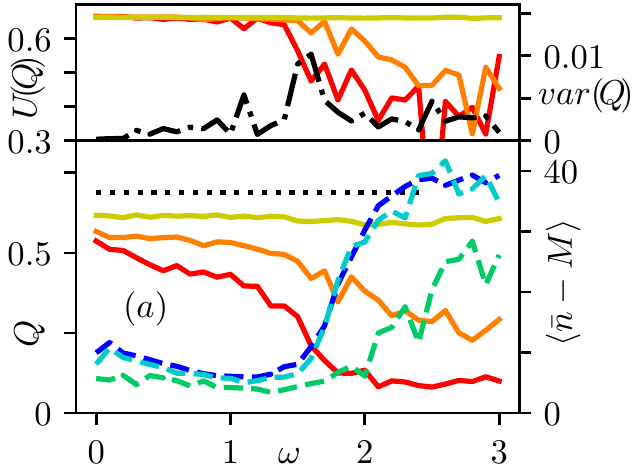}
    \includegraphics[width=0.3\textwidth]{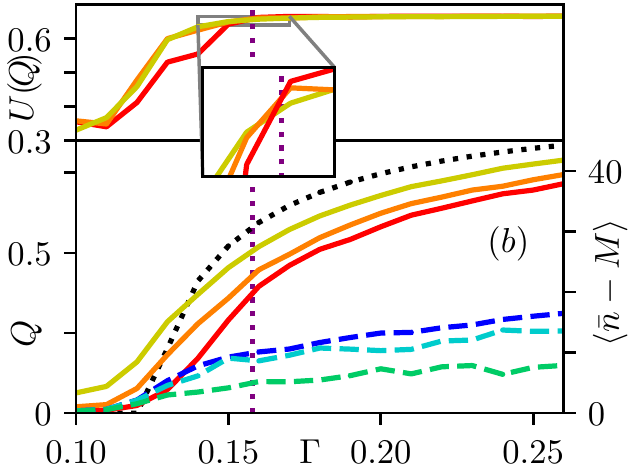}
    \includegraphics[width=0.3\textwidth]{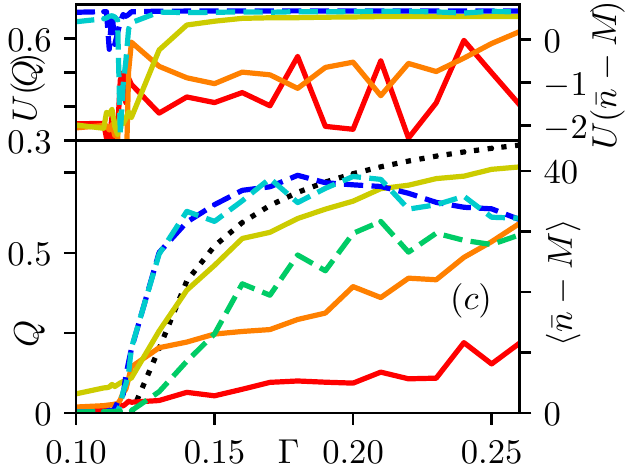}
\caption{Transitions between the phases $(a)$: $(ii)$ (homogeneous nematic) $\leftrightarrow$ $(iii)$ (nematic doplets) at $\Gamma=0.18$.
    The solid lines show the global nematic order parameter $Q$ (bottom) and its Binder cumulant (top) for $N=10^3$ (yellow), $N=10^4$ (orange) and $N=10^5$ particles (red).
    The dashed lines (bottom) shows the mean number of neighbors $\langle \bar{n} \rangle$ shifted by its mean field expectation value $M$ for $N=10^3$ (green), $N=10^4$ (turquoise) and $N=10^5$ (blue).
    The dash-dotted black line (top) shows the fluctuations of $Q$ for $N=10^5$,
    $(b)$: $(iv)$ (disorder) $\leftrightarrow$ $(ii)$ (homogeneous nematic) at $\omega=0.1$.
    Line styles and colors as in $(a)$.
    The black dotted line shows mean field results.
    The horizontal purple line shows an estimate of the $N\rightarrow \infty$ transition point obtained from the intersection of Binder cumulants. 
    $(c)$:  $(iv)$ (disorder) $\leftrightarrow$ $(iii)$ (nematic droplets) at $\omega=3$.
    Line styles and colors as in $(a)$.
    Dashed lines in top diagram show the Binder cumulant of the shifted mean number of neighbors for $N=10^4$ and $N=10^5$.
    All data are obtained from average of ten realizations.
    See SM \cite{SM} for numerical details.
    \label{fig:transitions_new}
    }
\end{figure*}
    
\subsection{Chiral}

For small to medium values of $\omega$ we observe a similar nematic order parameter as for $\omega=0$.
At small coupling there is disorder and above some critical value of $\Gamma$ there is global nematic order.
Compared to mean-field theory, the flocking transition is shifted towards higher coupling.
A similar effect was observed in the non-chiral system with polar alignment interactions \cite{KI21}, where the shift of the transition is caused by orientational correlations: a typical collision partner of a random particle is more likely to have a similar orientation.
Therefore, particles align stronger locally, but weaker globally, due to correlations.
It seems plausible that local correlations are enhanced by larger chirality because the effective diffusion coefficient reduces with increasing $\omega$.
Thus the same particles collide with each other for a longer time and therefore build up larger orientational correlations.
This argument seems to explain the shift of the transition towards higher coupling for larger $\omega$ as it is displayed in Fig.~\ref{fig:phasediagram_nemorder} $(a)$.

Within the nematically ordered phase $(ii)$ we do not see chiral phase separation like in the polar case, as predicted by mean field theory.
Instead, the nematically ordered phase is on large scales homogeneous, see Fig.~\ref{fig:snaps} $(ii)$.
On small scales particles accumulate into clusters, probably due to correlation effects, as this can not be seen in mean-field theory.
In Fig.~\ref{fig:transitions_new} $(b)$ we display the nematic order parameter $Q$, the Binder cumulant of $Q$ and the mean number of neighbors for three different system sizes as a function of coupling strength for the transition between phases $(iv)$ and $(ii)$.
The nematic order parameter indicates a continuous transition and we roughly estimate the position of the transition at infinite system sizes from the intersection of the Binder cumulants of different system sizes.
Within phase $(ii)$ the mean number of neighbors  moderately increases {with the coupling strength}, as we observe clustering at small scales as described above.

At large coupling and large chirality, we observe micro-phase separation of high density asynchronously nematic ordered droplets surrounded by a disordered low density gas, cf. Fig.~\ref{fig:snaps} $(iii)$, similarly to what it is observed in the polar case \cite{BennoPRL}.
This phase corresponds to the finite wavelength instability of the homogeneous flocking state that we discuss in Sec.~\ref{sec:meanfield}.
For large systems there is no global nematic order because the individual droplets are not synchronized.

In Fig.~\ref{fig:transitions_new} $(a)$ we display the transition between phases $(ii)$ and $(iii)$ that is characterized by a drop of the global nematic order parameter and a significant increase of the average number of neighbors at the same time.
The Binder cumulant of the nematic order parameter shows the characteristic shape of continuous transitions dropping from $2/3$ to smaller and smaller values for larger system sizes.
Here however, the data are too noisy to determine the infinite system size transition accurately.
Therefore, we also display the fluctuations of the nematic order parameter for the largest considered system size that shows a characteristic peak at about $\omega=1.6$, cf. top of Fig.~\ref{fig:transitions_new} $(a)$.

Eventually, we display the transition between phases $(iv)$ and $(iii)$ in Fig.~\ref{fig:transitions_new} $(c)$.
For the smallest considered system size it looks as there is a continuous transition from disorder towards nematic order.
However, as discussed above, for large system sizes there is no global nematic order also in phase $(iii)$, cf. Fig.~\ref{fig:transitions_new} $(c)$ (bottom).
Thus, the nematic order parameter can not be used to distinguish between phases $(iv)$ and $(iii)$.
However, we see in Fig.~\ref{fig:transitions_new} $(c)$ that there is a significant increase in the shifted mean number of neighbors $\langle\bar{n} -M\rangle$ indicating the transition from $(iv)$ to $(iii)$.
For large systems the Binder cumulant of the shifted mean number of neighbors shows a clear dip indicating a discontinuous transition.
For finite systems the transition is discontinuous because one (and for higher coupling more) high density droplet is formed.
However, the contribution of a single droplet to the global order parameter decreases with increasing system size.
Hence, the dip in the Binder cumulant is deeper for smaller systems and we expect the transition to be effectively continuous in the thermodynamic limit.

\section{Discussion and Conclusions\label{sec:summary}}
In summary, we model active chiral rods in two dimensions by means of point particles that move at constant speed and adopt their direction of motion according to an overdamped Langevin dynamics with nematic alignment interactions and inherent frequency $\omega$.
We study the phase behavior of the model by means of a kinetic mean field theory.
Depending on the coupling strength and independent of $\omega$ we find a continuous transition from disorder to nematic order under the assumption of spatially homogeneous distributions.
To clarify the validity of this assumption we study the linear stability of homogeneous solutions against  spatially extended perturbations.
The linear stability analysis goes beyond usual hydrodynamic treatments \cite{bertin2015comparison, BennoPRL} since all (infinite) angular modes of the homogeneous background solution and arbitrary many  modes of the perturbations are taken into account (although, in practice, one takes a finite number, here $k=-10, \dots, 10$).

We find no instabilities within the disordered phase $(iv)$.
In the nematically ordered phase, there is a long wavelength instability for $\omega=0$ that is confirmed in  simulations by the presence of macroscopic bands $(i)$ that are well known \cite{GinelliPeruani2010, Peruani16}.
For small to moderate chirality we find no instabilities of the homogeneous flocking state within mean field theory $(ii)$.
In simulations, we see small inhomogeneities due to correlation effects, however no strong clustering.
For larger chirality, there is a finite wavelength instability that manifests in the formation of asynchronous nematically ordered droplets surrounded by a disordered gas observed in simulations $(iii)$.
Simulation results show that the transitions $(ii) \leftrightarrow (iv)$ and $(ii)\leftrightarrow (iii)$ are continuous whereas the transition $(iii) \leftrightarrow (iv)$ is discontinuous.
The discontinuity in the $(iii)\leftrightarrow (iv)$ transition is due to the formation of a single nematically ordered droplet. 
Hence we expect the transition to become effectively continuous in the thermodynamic limit $N\rightarrow \infty$ because the impact of a single droplet on the order parameters decreases for larger system sizes. However further work is needed in order to conclude on this point. 

\section{Acknowledgments}
R.K. thanks Universitätsrechenzentrum Greifswald for supporting this work by providing computational resources.
and MICINN and the EU (Next Generation EU/PRTR) for funding through a ’María Zambrano’ fellowship. D. L. acknowledges MCIU/AEI/FEDER for financial support under Grant Agreement No.RTI2018-099032-J-I00.

\bibliographystyle{eplbib}

\newpage
\onecolumn
\section{Inhomogeneous Mean Field Theory\label{app:A}}
In this section we explicitly derive the space-dependent equivalent of Eq.~\eqref{eq:1FP}.
For that purpose we define a number of position dependent quantities.
First, we introduce the position dependent density
\begin{align}
    p_{\bold{r}}(\bold{r}):= \int_0^{2\pi} P_1(\bold{r}, \phi) d \phi.
\end{align}
Clearly, $p_\bold{r}$ is normalized
\begin{align}
    \int p_\bold{r}(\bold{r}) d \bold{r}=1,
\end{align}
thus it is a spatial probability density function.
In analogy to Eq. \eqref{eq:M} we introduce
\begin{align}
    M(\bold{r}):= N \int p_\bold{r}(\bold{r}+\tilde{\bold{r}}) \theta(R-|\tilde{\bold{r}}|)d \tilde{\bold{r}}.
\end{align}
Furthermore we introduce two local angular probability density functions as
\begin{align}
    p_\phi(\bold{r}, \phi):= \frac{P_1(\bold{r}, \phi)}{p_\bold{r}(\bold{r})},
    \qquad
    \bar{p}_\phi(\bold{r}, \phi)&:= \frac{N}{M(\bold{r})}\int P_1(\bold{r}+\tilde{\bold{r}}, \phi) \theta(R-|\tilde{\bold{r}}|)d \tilde{\bold{r}}.
    \label{eq:angular_space}
\end{align}
We can interpret these expressions as angular probability distributions where the spatial coordinate $\bold{r}$ is just an index.
The first expression in \eqref{eq:angular_space} specifies the distribution at position $\bold{r}$, whereas the second expression specifies the angular distribution averaged over a circle of radius $R$ around $\bold{r}$.
One easily checks that both distributions are normalized
\begin{align}
    \int_0^{2\pi} p_\phi(\bold{r}, \phi) d \phi=\int_0^{2\pi} \bar{p}_\phi(\bold{r}, \phi) d \phi=1.
\end{align}

For space-dependent distributions the equivalent of Eq. \eqref{eq:1FP} becomes with the above notation
\begin{align}
&\partial_t P_1(\bold{r}_1, \phi_1)=\partial_t [p_\bold{r}(\bold{r}_1)p_\phi(\bold{r}_1,\phi_1)]
\notag
\\
    &=-p_\bold{r}(\bold{r}_1)\partial_{\phi_1}\bigg\{M(\bold{r}_1) \Gamma \big[
    (\langle\sin(2\phi)\rangle_{\bar{p}_\phi(\bold{r}_1, \phi)} \cos(2\phi_1)    
    -\langle\cos(2\phi)\rangle_{\bar{p}_\phi(\bold{r}_1, \phi_1)} \sin(2\phi_1))\big]p_\phi(\bold{r}_1, \phi_1)\bigg\}
    \notag
    \\
    &\phantom{=} -\omega p_\bold{r}(\bold{r}_1)\partial_{\phi_1}p_\phi(\bold{r}_1,\phi_1)+D p_\bold{r}(\bold{r}_1)\partial_{\phi_1}^2p_\phi(\bold{r}_1,\phi)
    - v \cos(\phi_1)\partial_{x_1} p_\bold{r}(\bold{r}_1) p_\phi(\bold{r}_1,\phi_1) 
    - v \sin(\phi_1) \partial_{y_1} p_\bold{r}(\bold{r}_1) p_\phi(\bold{r}_1,\phi_1).
    \label{eq:1FP_space_dependent}
\end{align}
With Eq. \eqref{eq:1FP_space_dependent} we find a time evolution equation for the Fourier modes $F_{klm}(t)$ defined in Eq.~\eqref{eq:def_fourier}.
Neglecting quadratic terms in $F$ and assuming $\phi_0=0$ we obtain 
\begin{align}
    \partial_t F_{klm}=& \frac{\rho_0 \Gamma K_{l,m}}{I_0\bigg(\frac{M \Gamma Q}{2D}\bigg)}
    F_{2,l,m}\exp[i(2-k)\omega t]
    \bigg[\tilde{I}_{1-k/2}\bigg( \frac{M \Gamma Q}{2D}\bigg)
    + \frac{M\Gamma Q}{4D}\bigg\{ \tilde{I}_{2-k/2}\bigg( \frac{M \Gamma Q}{2D}\bigg)
    -\tilde{I}_{-k/2}\bigg( \frac{M \Gamma Q}{2D}\bigg)
    \bigg\}\bigg]
    \notag
    \\
    &+\frac{\rho_0 \Gamma K_{l,m}}{I_0\bigg(\frac{M \Gamma Q}{2D}\bigg)}
    F_{-2,l,m}\exp[i(-2-k)\omega t]
    \bigg[\tilde{I}_{-1-k/2}\bigg( \frac{M \Gamma Q}{2D}\bigg)
    + \frac{M\Gamma Q}{4D}\bigg\{ \tilde{I}_{-2-k/2}\bigg( \frac{M \Gamma Q}{2D}\bigg)
    -\tilde{I}_{-k/2}\bigg( \frac{M \Gamma Q}{2D}\bigg)
    \bigg\}\bigg]
    \notag
    \\
    &+F_{k-2,l,m}\frac{M\Gamma Q k}{2}[\cos(2\omega t)-i\sin(2\omega t)]
    +F_{k+2,l,m}\frac{M\Gamma Q k}{2}[-\cos(2\omega t)-i\sin(2\omega t)]
    -(i \omega k +D k^2)F_{klm}
    \notag
    \\
    &-v i \frac{2\pi}{L}l\frac{1}{2}[F_{k-1,l,m}+F_{k+1,l,m}]
    -v i \frac{2\pi}{L}m\frac{1}{2i}[F_{k-1,l,m}-F_{k+1,l,m}]
    ,
    \label{eq:time_evolution_modes}
\end{align}
where
\begin{align}
    K_{l,m}:&=\int_{-L/2}^{L/2}dx \int_{-L/2}^{L/2} dy \exp(il2\pi x/L)
    \exp(im2\pi y/L)\theta(R-|\bold{r}|)
    \notag
    \\
    &=\begin{cases}
    \pi R^2 \text{ if }m=l=0\\
    \frac{RL}{\sqrt{l^2+m^2}}J_1\bigg(\frac{2\pi}{L}R\sqrt{l^2+m^2}\bigg) \text{ else}
    \end{cases}
\end{align}
and
\begin{align}
    \tilde{I}_\nu(z):=\begin{cases}
    I_{|\nu|}(z) \text{ if }\nu \text{ is integer}\\
    0 \text{ else.}
    \end{cases}
\end{align}
Note that the linearized time evolution equation of perturbations from the homogeneous state, Eq.~\eqref{eq:time_evolution_modes} does not couple different spatial modes. Thus the analysis can be done for all values of the spatial wave vector $l, m$ separately.

In the numerical evaluation of the eigenvalues we considered angular wave vector components $k=-10, \dots, 10$.
We checked that the most unstable eigenvalues are left almost unchanged when taking into account more modes.
The spatially homogeneous long time solution \eqref{eq:angular_long_time} is not unique due to the arbitrary choice of $\phi_0$.
Thus, we expect that the spatial homogeneous modes, $l=0, m=0$ have an eigenmode with eigenvalue one, that corresponds to infinitesimal rotations of the homogeneous solution.
In practise, we find this eigenvalue to be slightly above one due to the discretization error introduced in Eq.~\eqref{eq:stability_matrix}.
We consider eigenmodes as unstable, if the absolute value of the eigenvalue is larger than the reference given by the the eigenvalue of infinitesimal rotations.

\section{Integration Scheme\label{sec:second_order_scheme}}

\begin{figure*}[h]
\centering
\includegraphics[width=0.32\textwidth]{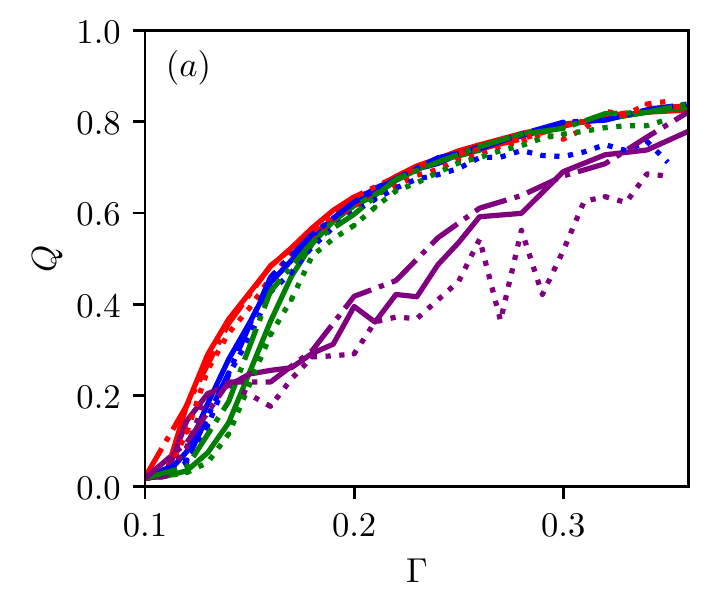}
\includegraphics[width=0.32\textwidth]{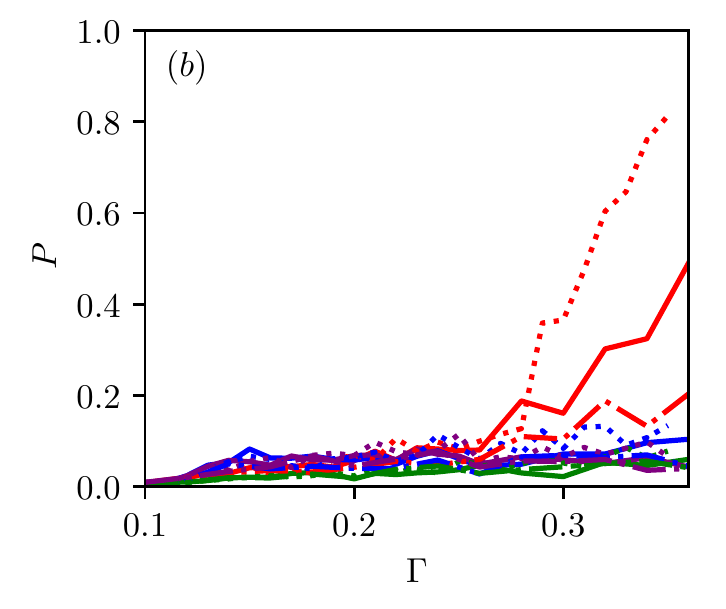}
\includegraphics[width=0.32\textwidth]{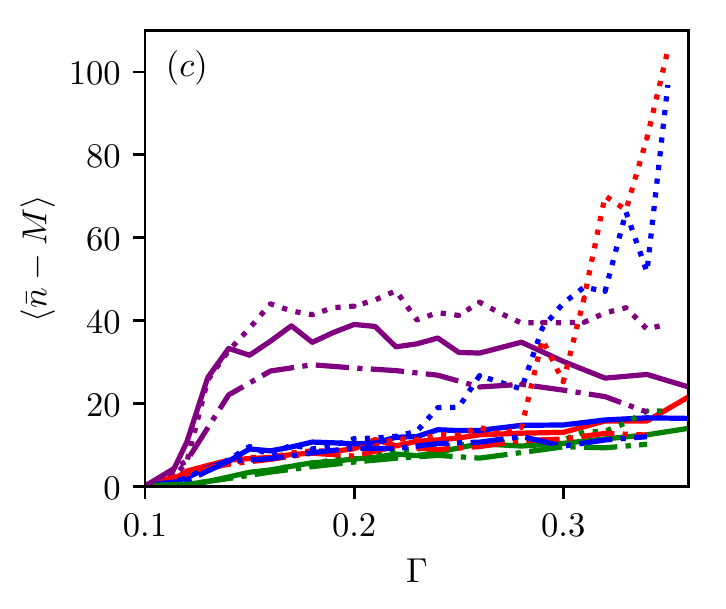}
\caption{Comparison between different integration schemes. Simulations for $N=10^4$, $\omega=0$ (red), $\omega=0.1$ (blue), $\omega=1.0$ (green) and $\omega=3.0$ (purple). We used an Euler-Maruyama scheme with $\Delta t=10^{-2}$ (dotted), $\Delta t=10^{-3}$ (dash-dotted) and a second order integration scheme with $\Delta t=10^{-2}$ (solid). We display the global nematic order parameter $(a)$, the global polar order parameter $(b)$ and the average number of neighboring particles $(c)$.
\label{fig:schemes}}
\end{figure*}
One of the simplest numerical discretization scheme of the stochastic differential equation \eqref{eq:model_langevin} is the well known Euler-Mayurama scheme
\begin{align}
    x_i(t+\Delta t)=& x_i+ \Delta t v \cos(\phi_i),
    \qquad
    y_i(t+\Delta t)= y_i+ \Delta t v \sin(\phi_i),
    \notag
    \\
    \phi_i(t+\Delta t)=& \phi_i+\Delta t\{\omega + \Gamma \sum_{j\in \partial_i}\sin[2(\phi_j-\phi_i)]\}
    +\sqrt{\Delta t}\sqrt{2D}\eta_i,
    \label{eq:euler_scheme}
\end{align}
where the time argument for all terms on the right hand side is $t$, which we omit for brevity. We denote the step size by $\Delta t$, and $\eta_i(t)$ are independent zero mean standard Gaussian random variables.

As an alternative we consider a weak second order scheme following \cite{KP92}.
For the considered system the scheme reads
\begin{align}
    x_i(t+\Delta t)=& x_i+\Delta t v \cos(\phi_i)-\frac{1}{4}\sigma^2 v \cos(\phi_i)(\Delta t)^2
    -\frac{1}{2} v \Gamma \sin(\phi_i)\sum_{j\in \partial_i}\sin[2(\phi_j-\phi_i)](\Delta t)^2
    \notag
    \\
    &- \frac{1}{2}v \sin(\phi_i)\omega (\Delta t)^2 - \frac{1}{2}(\Delta t)^{3/2}\sigma v \sin(\phi_i)\eta_i,
    \label{eq:second_order_scheme_x}
    \\
    y_i(t+\Delta t)=& y_i+\Delta t v \sin(\phi_i)-\frac{1}{4}\sigma^2 v \sin(\phi_i)(\Delta t)^2
    +\frac{1}{2}v \Gamma \cos(\phi_i)\sum_{j\in \partial_i}\sin[2(\phi_j-\phi_i)](\Delta t)^2
    \notag
    \\
    &+ \frac{1}{2}v \cos{\phi_i}\omega (\Delta t)^2 + \frac{1}{2}(\Delta t)^{3/2}\sigma v \cos(\phi_i)\eta_i,
        \label{eq:second_order_scheme_y}
    \\
    \phi_i(t+\Delta t)=&\phi_i  +\Delta t\omega + \Delta t\Gamma \sum_{j\in \partial_i}\sin[2(\phi_j-\phi_i)]
    - 2 \sigma^2 (\Delta t)^2\Gamma \sum_{j\in \partial_i}\sin[2(\phi_j-\phi_i)]
    \notag
    \\
    &+(\Delta t)^2\Gamma^2\sum_{j\in \partial_i}\big\{ \sum_{k\in \partial_j} \sin[2(\phi_k-\phi_j)]+\frac{\omega}{\Gamma}\big\}\cos[2(\phi_j-\phi_i)]
    -(\Delta t)^2\Gamma^2\big\{ \sum_{k\in \partial_i} \sin[2(\phi_k-\phi_i)]+\frac{\omega}{\Gamma}\big\}
    \notag
    \\
    &\times \sum_{j\in \partial_i}\cos[2(\phi_j-\phi_i)]
    +\sigma \sqrt{\Delta t}\eta_i + \sigma (\Delta t)^{3/2}\Gamma \sum_{j\in \partial_i}\cos[2(\phi_j-\phi_i)](\eta_j-\eta_i).
    \label{eq:second_order_scheme_phi}
\end{align}
Here, again we omitted the time variable $t$ for all terms on the right-hand side.
Note that the random variables $\eta_i$ that appear in all three equations are the same whenever index $i$ has the same value.
The second order scheme systematically considers  all terms up to order $(\Delta t)^2$.
Using cell lists, for homogeneous states the scheme \eqref{eq:euler_scheme} has complexity $N \cdot max(9 \rho,1)$ because for each particles one needs to find all neighbors (for which one needs to check all particles in a box of length $3$).
In principle, the second order scheme is of complexity $N\cdot  [max(9 \rho,1)]^2$ because one needs to find  all neighbors of neighbors of each particle.
In this study, we consider particularly high densities.
Furthermore, the effective density is even larger as soon as particles start to accumulate leading to a non-homogeneous density distribution.
Thus the second order scheme seems to be particularly inefficient and one might tend to use the Euler-Mayurama scheme with smaller time step instead.
However, due to the harmonic nature of the interactions between neighboring particles, we can simplify the second order scheme, Eq. \eqref{eq:second_order_scheme_phi}, such that it is of the same complexity as the Euler-Mayurama scheme.
Employing elementary trigonometric relations we obtain
\begin{align}
    \phi_i(t+\Delta t)=&\phi_i  +\Delta t\omega 
    + \Delta t\Gamma(1-2\sigma^2 \Delta t) \sum_{j\in \partial_i}\sin[2(\phi_j-\phi_i)]
    \notag
    \\
    &+\sum_{j\in \partial_i}\cos(2(\phi_j-\phi_i))\big\{ (\Delta t)^2\Gamma^2[S_j \cos(2\phi_j)-C_j \sin(2\phi_j)
    -S_i\cos(\phi_i)+C_i\sin(2\phi_i)] 
    \notag
    \\
    &+\sigma (\Delta t)^{3/2} \Gamma (\eta_j-\eta_i)\big\}
    +\sigma \sqrt{\Delta t}\eta_i,
    \label{eq:so_simplyfied}
\end{align}
where
\begin{align}
    S_i:=\sum_{k\in \partial_i}\sin(2 \phi_k),
    \qquad
    C_i:=\sum_{k\in \partial_i}\cos(2 \phi_k).
    \label{eq:SC}
\end{align}
Practically, in a first loop over all particles and particles neighbors we calculate the vectors $S_i$, $C_i$ and also choose the noise realization $\eta_i$ and in a second loop over all particles we update positions according to Eqs. \eqref{eq:second_order_scheme_x}, \eqref{eq:second_order_scheme_y} and angles according to \eqref{eq:so_simplyfied}.

We compare simulations of both schemes using step sizes $\Delta t=10^{-2}$ (method 1) and $\Delta t=10^{-3}$ (method 2) for the Euler-Mayurama scheme and $\Delta t=10^{-2}$ (method 3) for the second order scheme. 
In Fig.~\ref{fig:schemes} we compare global nematic order parameter $Q$, global polar order parameter $P$ and the shifted mean number of neighbors $\langle \bar{n}-M\rangle$ measured for $N=10^4$ for the different schemes at different values of $\omega$.

We do not see a big impact of the scheme or chosen time step on the global nematic order parameter, cf. Fig.~\ref{fig:schemes} $(a)$.
Nevertheless, in particular for $\Delta t=10^{-2}$, we observe artifacts at large coupling.
This can be seen best in the polar order parameter for $\omega=0$, cf. Fig.~\ref{fig:schemes} $(b)$.
In reality, or when using a smaller time step, there is no polar order in this model.
However, if the time step is chosen too large it seems as there would be polar order.
This apparent polar order goes along with the formation of huge clusters as we can see from the average number of neighbors, cf. Fig.~\ref{fig:schemes} $(c)$ for $\Delta t=10^{-2}$, method 1, for $\omega=0, 0.1$.
We believe that discretization errors become significant when too many particles accumulate locally due to the additive nature of the interaction \eqref{eq:model_langevin}.
Apparently this effect is self-amplifying because within the inaccurate scheme particles accumulate even stronger.

Choosing either smaller time steps (method 2) or a second order integration scheme (method 3) appears to avoid those undesired artifacts.
With method 3 we completely avoid the strong particle accumulation, see Fig.~\ref{fig:schemes}
 $(c)$ and the artificial polar flocks are seen only at higher couplings compared to method 1, see Fig.~\ref{fig:schemes} $(b)$.
 Method 2 performs even better than method 3, however, it is computationally more expensive.
 Therefore, we used only method 3 for all simulations of the letter and considered not too large couplings, $\Gamma \le 0.26$, such that there are no simulation artifacts present.
 
 In summary, it should be pointed out that special care has to be taken when choosing integration scheme and time step for models with additive interactions as we consider here \eqref{eq:model_langevin}.
 For too large time steps, unphysical artifacts can be observed.
 This can be avoided using higher order integration schemes or smaller time steps.
 Due to the harmonic nature of the interaction \eqref{eq:model_langevin}, second (and in principle also higher) order schemes run at the same complexity as the Euler-Mayurama scheme.
 Therefore, higher order schemes appear to be superior for the considered type of interactions.

 \section{Numerical Details} 
 Fig. \ref{fig:snaps}: physical parameters: $R,v,D=1$, $N=10^5$, $L_x,L_y=100$, integration scheme: method 3 with $\Delta t=10^{-2}$, initial conditions: random, uniform and isotropic, thermalization: $1009999$ time steps.
 Fig. \ref{fig:phasediagram_nemorder} $(a)$: physical parameters: $R,v,D=1$, $N=10^5$, $L_x,L_y=100$, integration scheme: method 3 with $\Delta t=10^{-2}$, initial conditions: random, uniform and isotropic, thermalization: $999999$ time steps, measurement averaged over $10$ realizations and $10^4$ time steps for each realization.
 Fig. \ref{fig:transitions_new}: physical parameters: $R,v,D=1$, $N=10^3, 10^4, 10^5$, $L_x,L_y=10, \sqrt{10}\times 10, 100$, integration scheme: method 3 with $\Delta t=10^{-2}$, initial conditions: random, uniform and isotropic, thermalization: $999999$ time steps, measurement averaged over $10$ realizations and $10^4$ time steps for each realization.
 Simulation data of method 3 are available at \cite{data1}, simulation data of methods 1, 2 at \cite{data2}.

\end{document}